\ifx\pdfoutput\undefined
  \documentclass{aastex}
  \usepackage{natbib}
  \usepackage{emulateapj5}
  \usepackage{graphicx}
  \DeclareGraphicsExtensions{.ps, .jpg}
\else
  \documentclass[pdftex]{aastex}
  \usepackage{natbib}
  \usepackage{emulateapj5}
  \usepackage{graphicx}
  \usepackage[colorlinks,hyperindex]{hyperref}
\fi

\newcommand{\ii}{$I_{814W}$}

\begin{document}

\bibliographystyle{apj}

\submitted{Draft version \today}

\title{The Visibility of Galactic Bars and Spiral Structure At High Redshifts}

\author{Sidney van den Bergh}
\affil{
   Dominion Astrophysical Observatory,
   Herzberg Institute of Astrophysics,
   National Research Council,
   5071 West Saanich Road,
   Victoria, British Columbia,
   V9E 2E7, Canada.\\
   sidney.vandenbergh@nrc.ca}

\medskip

\author{Roberto G. Abraham}
\affil{
   Department of Astronomy \& Astrophysics,
   University of Toronto, 60 St. George Street,
   Toronto, ON, M5S 3H8, Canada.\\
   abraham@astro.utoronto.ca}

\medskip

\author{Laura F. Whyte, Michael R. Merrifield}
\affil{
   School of Physics \& Astronomy,
   University of Nottingham,
   Nottingham NG7 2RD, UK. \\
   ppxlfw@unix.ccc.nottingham.ac.uk, michael.merrifield@nottingham.ac.uk}

\medskip

\author{Paul Eskridge}
\affil{
Department of Physics \& Astronomy, Minnesota State University
Mankato, MN   56001.\\ paul.eskridge@mnsu.edu
}

\medskip

\author{Jay A. Frogel, Richard Pogge}
\affil{
   Department of Astronomy,
   Ohio State University,
   140W 18th Avenue, Columbus, OH,
   43210.\\
   frogel@astronomy.ohio-state.edu,
pogge@astronomy.ohio-state.edu, eskridge@astronomy.ohio-state.edu}

\begin{abstract}
We investigate the visibility of galactic bars and spiral
structure in the distant Universe by artificially redshifting 101
$B$-band CCD images of local spiral galaxies from the Ohio State
University {\em Bright Spiral Galaxy Survey}. These local galaxy
images represent a much fairer statistical baseline than the
galaxy atlas images from Frei et al. (1995), the most commonly
used calibration sample for morphological work at high redshifts.
Our artificially redshifted images correspond to Hubble Space
Telescope $I_{814}$-band observations of the local galaxy sample
seen at $z=0.7$, with integration times matching those of both the
very deep Northern Hubble Deep Field data, and the much shallower
Flanking Field observations. The expected visibility of galactic
bars is probed in two ways: (1) using traditional visual
classification, and (2) by charting the changing shape of the
galaxy distribution in ``Hubble space'', a quantitative
two-parameter description of galactic structure that maps closely
on to Hubble's original tuning fork. Both analyses suggest that
over 2/3 of strongly barred luminous local spirals ({\em i.e.}
objects classified as SB in the {\em Third Reference Catalog})
would still be classified as strongly barred at $z=0.7$ in the
Hubble Deep Field data. Under the same conditions, most weakly
barred spirals (classified SAB in the {\em Third Reference
Catalog}) would be classified as regular spirals.  The
corresponding visibility of spiral structure is assessed visually,
by comparing luminosity classifications for the artificially
redshifted sample with the corresponding luminosity
classifications from the {\em Revised Shapley Ames Catalog}. We
find that for exposures times similar to that of the Hubble Deep
Field spiral structure should be detectable in most luminous
($M_B \sim M^\star$) low-inclination spiral galaxies at $z=0.7$ in
which it is present. However, obvious spiral structure is only detectable
in $\sim 30$\% of comparable galaxies in the HDF Flanking Field
data using WF/PC2. Our study of artificially redshifted local galaxy
images suggests that, when viewed at similar 
resolution, noise level and redshift-corrected wavelength, 
barred spirals are less common
at z $\sim0.7$ than they are at z = 0.0, although more data
are needed to definitively rule out the possibility that
cosmic variance is responsible for much of this effect.
\end{abstract}


\section{INTRODUCTION}
\label{sec:introduction}

The {\em Hubble Space Telescope} has, for the first time, allowed
the direct study of galaxies as they appeared when the Universe
was much younger than it is at the present time. Inspection of the
images of such distant galaxies shows that the Hubble tuning fork
diagram does not provide an adequate framework for the
classification of distant  galaxies (van den Bergh et al. 1996)
and that the fraction of peculiar and merging galaxies increases
with redshift (Abraham et al. 1996). Around $1/3$ of luminous
galaxies lie off Hubble's tuning fork at $z=1$ (Brinchmann et al.
1998). A number of studies have also claimed that the percentage
of barred spirals seems to decrease towards larger look-back times
(van den Bergh et al. 1996; Abraham et al. 1999). This trend has
been based on the results from objective, but rather
coarse-grained, automated morphological classifications. More
fine-grained ``precision morphology" (which can subdivide spiral
galaxies into various classes, for example) still requires a
trained human eye. On this basis van den Bergh et al. (2000) and
van den Bergh, Cohen \& Crabbe (2001) note that the fraction of
``grand design" spirals appears to decrease rapidly with
increasing redshift. It is important to emphasize that these studies
focus mainly on galaxies
at redshifts $z<1$, so that systematic changes in galaxy appearance 
due to shifting of the filter bandpass in the rest frame of the galaxy are
easily understood. For example, at $z\sim0.7$, HST imaging using the F814W filter
corresponds quite closely to $B$-band imaging in the rest frame of the galaxy.

However, an alternative interpretation of the absence of barred
and grand design spirals at high redshifts is that the features
that define these categories of objects locally are undetectable
outside of the nearby Universe. The visibility of these features
at high redshifts has been poorly understood because of the
absence of a suitable CCD imaging sample of representative local
galaxies. Such samples can be used to test the visibility of fine
structures in galaxies by ``artificial redshifting'', in which
images are binned and noise added in order to mimic the appearance
of high-redshift counterparts. In the present paper we will adopt
this strategy using a new sample of local galaxy images taken from the {\em Ohio State University
Bright Spiral Galaxy Survey} (Frogel, Quillen \& Pogge 1996). This
sample provides a more statistically fair representation of the
local morphological mix than other samples used to calibrate
high-redshift morphology by artificial redshifting. 

In the present paper we quantify the systematic effects of
decreasing resolution (and of increasing noise) on the visibility
of galactic bars and grand design spiral structure using a mixture
of techniques. Where possible, we will back up traditional visual
classifications using quantitative measures. A useful tool in this
regard is what Abraham \& Merrifield (2000) have dubbed ``Hubble
space'', a quantitative two-parameter description of galactic
structure that maps closely on to Hubble's original tuning fork.
When no suitable quantitative measure exists, as for describing
the existence of grand design spiral structure, we
unapologetically adopt a purely visual approach.

A plan for this paper follows. In \S\ref{sec:osu} we describe our
sample, and highlight its advantages (and deficiencies) when it is
used to calibrate morphology at intermediate redshifts of around
$z=0.7$. In \S\ref{sec:viz} we revisit the visual classifications
for the galaxies in the sample, and compare the classifications
used in the present paper with those from existing catalogs. In
\S\ref{sec:correlations} we introduce the Hubble space diagram for
our sample. In \S\ref{sec:comparison} we describe our artificially
redshifted sample, and analyze the robustness of various features
in the galaxy images using both visual classifications and
quantitative measures. The visibility of grand design spiral
structure is analyzed in \S\ref{sec:grand}. Our results are
discussed in \S\ref{sec:discussion}, and our conclusions
summarized in \S\ref{sec:conc}. Finally, we note that throughout
the present paper we adopt a cosmology where $h=0.7$,
$\Omega_M=0.3$, and $\Omega_\Lambda=0.7$.

\section{THE OSU BRIGHT SPIRAL GALAXY SURVEY}
\label{sec:osu}

Our data set consists of 101 $B$-band images 
of galaxies taken from the Ohio
State University {\em Bright Spiral Galaxy Survey} (BSGS;
Frogel, Quillen, \& Pogge 1996; Eskridge et al., in preparation). The BSGS is a survey consisting of
$B, V, R, J, H, K$ images for 205 objects selected from
the {\em Third Reference Catalog of Bright Galaxies} (RC3;
de Vaucouleurs et al. 1991).  Our sub-sample of these data
consists of all $B$-band images that were
available at the time of writing. The BSGS is constructed from the
RC3 on the basis of three simple criteria: (1) no E or S0 galaxies
are included; (2) $B < 12$ mag; (3) galaxy diameter $\le$
6.5\arcmin. The reader is referred to Frogel, Quillen \& Pogge (1996) for
details. The diameter limit imposed on the
sample was intended to ensure that all galaxies fit on the various detectors used,
and is relatively unimportant as 95\% of the $B<12$ mag spiral
galaxies in the RC3 are imaged in the BSGS.

\begin{figure*}[htb]
    \centering
    \includegraphics[width=4.5in]{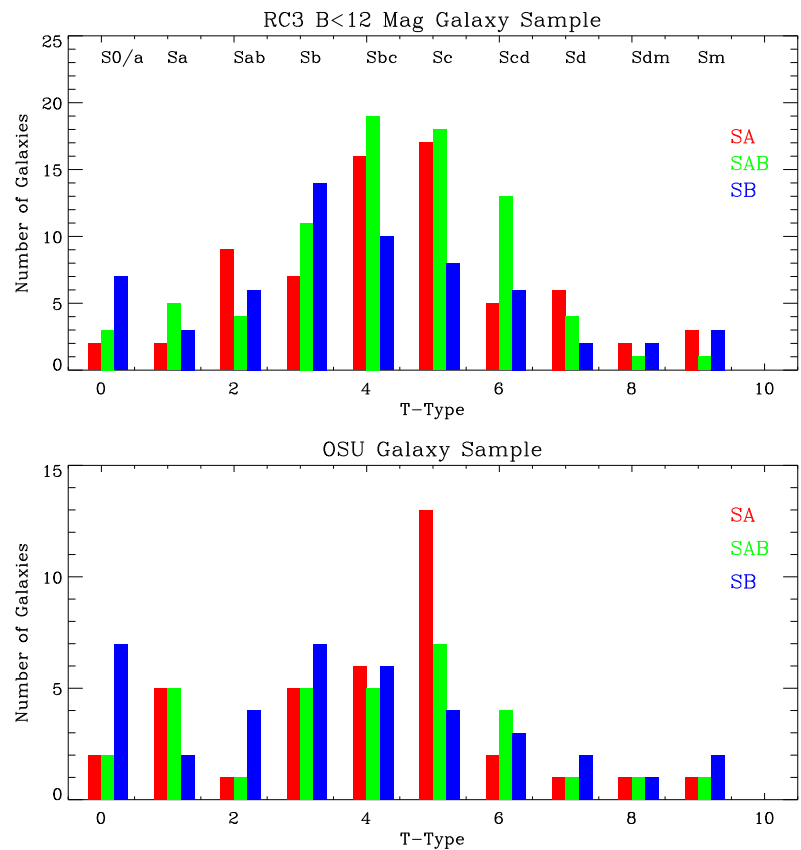}
    \caption{An illustration of the statistical fairness of the OSU
    Bright Spiral Galaxy Survey data for use in morphological investigations.
    Distributions of unbarred [red], weakly barred [green], and
    strongly barred [blue] galaxies are plotted as a function of
    Hubble stage. The upper panel shows the distribution for the
    Third Reference Catalog (RC3), while the lower panel shows the
    corresponding distribution for the present galaxy sample.}
    \label{fig:rc3comparison}
\end{figure*}

It is important to consider whether any important biases are
introduced by restricting our consideration to a subset of
the full BSGS in this manner (we re-emphasize that since the BSGS survey as a whole is
95\% complete, the full survey has no important biases with respect to the RC3 selection
criteria). In Figure~\ref{fig:rc3comparison} we
present histograms showing the frequency of the barred and
regular spirals in our sample of 101 galaxies from the BSGS as a
function of Hubble stage, and compare
these with the corresponding histograms for the full RC3 cut
at the same limiting magnitude as the BSGS. Note that in this
figure (and throughout the remainder of this paper), we will adopt the
terminology of the RC3 which denotes strongly barred spirals as
class SB (shown in blue), weakly/tentatively barred systems as
class SAB (shown in green), and unbarred spirals as class SA
(shown in red)\footnote{Note that in the standard encoding from
the RC3 given column 2 of Table~1, SAB spirals are encoded as type
SX.}. A Kolmogorov-Smirnov test confirms the visual impression
from this figure that the morphological mix of galaxies in our
BSGS sub-sample is statistically similar to that of the full RC3
cut at $B<12$ mag. (The Kolmogorov-Smirnov test returns a formal
33\% probability that the AB or B distributions differ, and a 60\%
probability that the A galaxy distributions differ).

It is interesting to compare Figure~\ref{fig:rc3comparison} with
the corresponding histograms for the sample of Frei et al. (1996),
which has been used by many authors (including several of the
co-authors of this paper) to calibrate the expected appearance of
high-redshift galaxies. These are shown in Figure~1 of Abraham \&
Merrifield (2000). The Frei et al.\ (1996) sample is quite
strongly biased against both Sa spirals and very late-type spirals
(Sd and beyond); near the endpoints of the tuning fork, some early
and late type spirals are included in the Frei et al.\ (1996)
sample, but with some bar classes missing. For example,
strongly barred SBab galaxies are absent from the sample, while
all Scd galaxies in the sample are weakly barred SABcd systems. In
contrast, the present BSGS sample is a much less biased representation
of the local morphological mix for luminous galaxies.

While we consider the present local galaxy CCD imaging sample a
major step forward, it is important to note that in some ways this
sample is still far from ideal for use in calibrating galaxy
morphology at $z\sim 0.7$, as indeed are all magnitude-limited
samples of bright galaxies. This is clear from
Figure~\ref{fig:absmag} which shows the $B$-band absolute
magnitude distribution for galaxies in the BSGS, along with a
Schechter function empirically normalized to the bright shoulder
of the distribution. As expected from a magnitude-limited sample, the
absolute magnitude distribution for galaxies in BSGS is peaked at
$M_\star$ and drops off sharply at both the bright and faint ends.
At $z=0.7$ the \ii~ filter on HST is roughly synchronized to
rest-frame $B$-band, so the effects of ``morphological
K-corrections'' are small, and $B$-band local images are
well-matched to \ii-band HST data. Therefore, on
figure~\ref{fig:absmag} we also show dashed lines which indicate
the absolute magnitude limits corresponding to $I_{814}=22$ mag,
the magnitude limit adopted for morphological classification in
the {\em Medium Deep Survey} (Griffiths et al. 1994; Glazebrook et
al. 1995; Abraham et al. 1996), and $I_{814}=23.2$ mag, the
magnitude limit corresponding to the bar visibility study of
Abraham et al. (1999). On the basis of this figure, it is clear
that: (i) the BSGS is reasonably well-matched to shallow (several
orbit) HST imaging observations, such as those of the Medium Deep
Survey (MDS), the Hawaii Deep Survey (HDS), and the CFRS/LDSS
imaging surveys, but (ii) deep HST observations, such as those in
the Hubble Deep Fields, will probe much further down the
luminosity function than the galaxies represented in the BSGS.
This is true even in the no-evolution case, but ultimately how far
down the LF we look depends critically on the amount of luminosity
evolution. With strong evolution we can of course probe distant
galaxies that would have been quite faint in the absence of
luminosity evolution. The implications of this will be considered
in Section~\ref{sec:discussion}.

\section{VISUAL CLASSIFICATIONS}
\label{sec:viz}

In order to allow a consistent comparison between the artificially
redshifted morphologies of the galaxies in our sample and their
local morphologies, one of us (vdB) visually reclassified the
local galaxy sample using the DDO system of van den Bergh
(1960abc). Results from this procedure are shown in Table~1 (along with the
classifications obtained from repeating this exercise with artificially redshifted and noise-degraded
galaxy images, the results from which will be described in \S\ref{sec:comparison}).
Hubble types are available for 91 of the galaxies in our sample
from the {\em Carnegie Atlas of Galaxies} (Sandage \& Bedke 1994).
A comparison between our classifications and those in the {\em
Carnegie Atlas} shows excellent agreement. The standard deviation
of the differences between the Hubble types in these two data sets
is (after uncertain classifications, marked in Table 1 by a colon,
have been excluded) 0.41 Hubble class.  The mean difference
between Hubble classification types for these two sets of data is
found to be 0.00 +/- 0.04 Hubble class. The overall fraction of
barred galaxies in both samples are also in good agreement, as
shown in Table~2\footnote{Note that Sandage does not use de
Vaucouleurs' notation SAB for weak bars, but uses the designation
S/SB for objects of intermediate type. In Table~2 the amalgamation
of vdB types S(B) and SB should be compared with Sandage's numbers
for type SB}. We note that the fraction of barred galaxies in the RC3 is in good agreement
with the corresponding value determined independently for the BSGS by Eskridge et al. (2000).

Gratifyingly good agreement is also found between the present luminosity
classifications of galaxies and those by Sandage \& Tammann (1981).
These are also listed in Table 1. (Note that the quoted Carnegie
classifications are simplified: outer rings are not marked, ring
(r) and spiral (s) subclasses are omitted, and luminosity classes
are simplified, i.e. Sandage \& Tammann's class I.7 is quoted as
I-II). After excluding objects with uncertain classifications, 43
pairs of luminosity classifications were available. The standard
deviation of the difference between these classifications was found
to be $0.66$ luminosity classes. The mean difference, in the sense
van den Bergh minus Carnegie, is $+0.10 \pm 0.05$ luminosity
classes, indicating that the Sandage \& Tammann (1981) luminosity
classes are very close to the DDO system of van den Bergh
(1960abc).

\section{CORRELATIONS IN LOCAL HUBBLE SPACE}
\label{sec:correlations}

Using the sample of Frei et al. (1996), Abraham \& Merrifield
(2000) defined a quantitative two-dimensional morphological
parameterization (``Hubble space'') whose $x$-coordinate measures
central concentration of light, and whose $y$-coordinate measures
the degree to which a galaxy is barred in a quantitative way. A
remarkably large amount of information concerning the properties
of local galaxies can be inferred from a close inspection of
Hubble space (Whyte et al., in preparation), but for our present purposes its
most important characteristic is that it provides
a convenient benchmark allowing us to understand how the
systematics of galaxy morphology vary as a function of resolution
and signal-to-noise ratio. In this section we will
describe the basic features of the local galaxy distribution in
Hubble space, before moving on to describe how these properties
change when our galaxy sample is artificially redshifted to z=0.7.

The distribution of the present BSGS sample in Hubble space is
shown in Figure~\ref{fig:local}. In the top-left panel of this
figure, colored symbols keyed to the scheme used in
Figure~\ref{fig:rc3comparison} are used to subdivide the galaxy
population into bar types, based on classifications from the RC3.
The dashed line shown in the figure subdivides barred from
unbarred galaxies extremely cleanly. As was found for the Frei et
al. (1996) sample analyzed in Abraham \& Merrifield (2000), there
is essentially no intermixing of unbarred galaxies and strongly
barred galaxies in Hubble space, and weakly barred spirals are
seen to be predominantly late-type (ie. low central
concentration). However, unlike the corresponding figure for the
Frei et al. (1996) sample no obvious ``tuning fork'' shape (or
even bimodality) emerges from this diagram.

In the top right panel of Figure~\ref{fig:local} we show the same
galaxies distribution in Hubble space, except with the symbols
keyed to Hubble stage. This panel shows
the obvious gradient in Hubble stage as a function of central
concentration rather strikingly. This illustrates why Hubble space is a
successful representation of the local galaxy data --- measures of
central concentration are clearly rather closely linked to Hubble
stage. However, this panel provides a warning
against basing morphological classifications of distant galaxy
images {\em solely} on perceived central concentration or
bulge-to-disk ratio (Abraham 1999). The mixed criteria
(bulge-to-disk ratio, tightness of spiral structure, and degree of
resolution in the arms) used to define visual classifications on
the classical tuning fork make distinguishing between various
types later than T=5 impossible on the basis of central
concentration of light. This is in accord with the general ``rule
of thumb'' that central concentration is only used as a
classification criterion for objects of early and
intermediate-type. Late types are generally seen with open enough
arms that an adequate classification can be made without invoking
any visual bulge-to-disk ratio criterion.

The distribution of galaxies in Hubble space keyed to rotational
asymmetry (determined using the method given in Abraham et al. 1996)
is shown in the lower left panel of
Figure~\ref{fig:local}. It is interesting that SB galaxies appear
to be made up of two populations: a tight sequence of symmetric
objects and an outlier population of highly asymmetric galaxies.
Furthermore, a {\em hint} of bimodality (whose reality will be
interesting to test once the full BSGS data becomes available)
appears in the plot when consideration is restricted to galaxies
with a high degree of symmetry. Our tentative explanation for the tuning fork
shape seen in the Hubble space distributions of Abraham \&
Merrifield (2000) is that the galaxies in the Frei et al. (1996)
sample were chosen to be visually ``pretty'', and it seems this
may be correlated with low asymmetry. In this connection it is
interesting to compare measurements of rotational asymmetry with
more traditional visual methods for characterizing textural
aspects of galaxy structure.  In the lower right hand panel of
Figure~\ref{fig:local} we show BSGS Hubble space distribution
keyed to luminosity classification.  For convenience, luminosity
classes have been mapped onto a numerical sequence (eg. type I =
1.0, II-III = 2.5, etc.). Galaxies with no luminosity
classifications are shown as small black circles.

Most spirals in our sample are of luminosity class II. Because
rotational asymmetry is measured by summing pixel-by-pixel
differences, ``ragged'' spiral structure might be expected to
slightly increase measured rotational asymmetry, although the
effect is probably very small. In this connection it is
interesting to compare measures of asymmetry and luminosity
classification. A comparison between the lower two panels shows
that luminosity and asymmetry are only very loosely correlated.
Note however that luminosity classifications can only be applied
to a subset of all systems in our sample. This is mainly because
luminosity classifications are not defined for early-type spirals,
because the spiral features in these objects are too short.
Another factor is more subtle. For luminosity classes I-III  most
of the weight in the classification comes from the strength and
structure of the spiral arms. However, in classes III-V most of
the weight comes from surface brightness, with the dwarfs having
the lowest surface brightness. On Palomar Sky Survey plates all
images have similar exposure times so that the apparent surface
brightness is related to the intrinsic surface brightness, and
consistent luminosity classifications may be assigned to bulk
samples of galaxies. However, this is {\em not} the case for the
present images. As a result it is, on the basis of our present
images, very difficult to assign luminosity classes to spirals of
classes III-V. In practice this is not be much of a problem
because most spirals in the our sample are intrinsically luminous.

\section{ARTIFICIALLY REDSHIFTED IMAGES}
\label{sec:comparison}

Simulated z=0.7 \ii-band HST images (corresponding to
rest-frame $B$-band) for the 101 galaxies in our
sample were constructed using the techniques described in Abraham
et al. (1996). Two sets of synthetic images were constructed,
mimicking the exposure time and sampling of the \ii-band image of
the Northern Hubble Deep Field (123600s exposure with
0.04\arcsec/pixel), and of a typical flanking field image (4000s
exposure, 0.1\arcsec/pixel). Parameters defining Hubble space were
re-measured from these degraded frames, which were also visually
classified onto the DDO system by vdB. The images were given a
final visual inspection by one of us (RGA), who assigned a simple
numerical index to each image denoting the visibility of spiral
structure, as described in \S\ref{sec:grand}. For the synthetic
Hubble Deep Field images, visual inspections were limited to
\ii$<25$ mag, corresponding to the magnitude limit typically used
by investigators working on galaxy morphology in the HDF.
Similarly, visual inspection of the synthetic flanking field data
was limited to \ii$<22$ mag. A summary of our results for the
simulated HDF and FF data is included in Table~1. Classifications
in the simulated Flanking Field images have been limited to $I<22$
mag (corresponding to the typical magnitude limit from the Medium
Deep Survey data).

\subsection{Robustness of Visual Classifications}
\label{subsec:visual}

Results from the visual classification of the present sample are tabulated in
Table~2, which shows a comparison between our local bar strength
classifications and those at $z = 0.7$ for both the HDF
signal-to-noise ratio and for the FF signal-to-noise. The table
shows the following: (1) The fraction of SB + S(B) galaxies is
28\% for both the Sandage \& Tammann (1981) classifications in the
Carnegie Atlas and in van den Bergh's classifications of the BSGS
galaxies at $z = 0.0$. If the resolution of images is degraded to
that of Hubble Deep Field galaxies at $z = 0.7$ then the fraction
of SB + S(B) galaxies that can still be recognized as such drops
from 28\% to 19\%. (3) At first sight it might be unexpected that
the fraction of SB + S(B) galaxies that can still be recognized as
such remains constant at 19\% when the signal-to-noise ratio is
decreased to that in the Flanking Fields, The reason for this is
probably that the FF galaxies ,which all have $I_{814} < 22.0$ are
brighter than the HDF galaxies, for which $I_{814} < 23.2$. In
summary the main result obtained from Table 2 is that roughly 2/3
of strongly barred galaxies will still be recognizable as such at
z = 0.7. For weakly barred galaxies the fraction of
recognizable bars drops from 17\% at $z=0.0$ to $\sim12$\% at $z =
0.7$ for HDF S/N and $\sim10$\% for FF S/N.

Among the 101 local spiral galaxies with images degraded to z =
0.7 that are listed in Table 1, 81 are of type S, 12 of type S(B)
S(B:) or S(B?) and are 8 of type SB. For comparison van den Bergh,
et al. (2000) found only S galaxies and no objects of types S(B)
or SB  among 18 comparable HDF spirals with $0.55 < z < 0.85$.
Taken at face value this result appears to confirm the conclusion
by van den Bergh et al. (1996) that barred spirals are deficient
at large redshifts. Among the local spirals listed in Table 2, a
total of 63 have $I < 22$ and may therefore be compared to those
in the Flanking Fields for which van den Bergh et al. (2000) have
given classifications. Of the galaxies in Table 2 that have $I <
22$, 51 are found to be of type S, 7 of type S(B) and 5 of type
SB. For comparison the numbers of comparable spirals in the
Flanking Fields with $0.55 < z < 0.85$ are found to be  42 of type
S, 2 of type S(B) and 0 of type SB. It is concluded that at
comparable resolution and noise the fraction of barred spirals in
the Flanking Fields (2/44 = 5\%) appears to be lower than it is
among nearby galaxies (12/63 = 19\%).

A graphical summary of the results presented in this section is shown in Figure \ref{fig:secsummary}.  This figure also sheds some light on the statistical significance of claims for an absence of bars in the Hubble Deep Field observations. The relative bar fractions determined from the artificially redshifted and observed samples suggests that the absence of high-redshift bars is formally at least a 2$\sigma$ effect. However, as described in the \S\ref{sec:discussion}, it is wise to remain cautious given the small volume of the Hubble Deep Fields, and the possibility of environmental effects that {\em might} skew the relative abundance of barred spirals as a function of environment. 

Before leaving visual classifications (which we will return to in \S\ref{sec:grand}) to consider their automated counterparts in the next section, it is worth emphasising a final point: when spiral structure is not visible in a high-redshift galaxy, the system is {\em not} usually classified as a peculiar galaxy on this basis by experienced morphologists (although a local disk galaxy without spiral arms would probably be classed as peculiar, or at least anemic). This is because spiral structure washes away at low signal-to-noise levels while the nuclear bulge and disk do not.  Therefore such systems are still classified as a spiral galaxies (in spite of the absence of spiral structure) whose Hubble stage is determined solely on the basis of apparent bulge-to-disk ratio. The ramifications of this for both visual and automated classification  are described in Abraham (1999).  It is also worth emphasizing that in classification of galaxies on deep HST data, such as the the Hubble Deep Field, van den Bergh denotes objects as ``peculiar'' only when they are not recognizable as disk or spheroid systems at all (e.g. when they have multiple nuclei and gross distortions). In most cases a recognizable disk exhibiting peculiarity does so at a level where the distortion is best viewed as a perturbation superposed on a well-known galaxy form.  Such systems are {\em sub-classed} as peculiar, not {\em super-classed} as peculiar. For example, an early-type disk galaxy with a warp might be classed ``Sa (pec)", but not as ``pec''. The distinction between an object sub-classed as peculiar, and an object so peculiar that no association with a disk or spheroid is possible, is somewhat subjective at present. It would be worthwhile for this distinction to be placed on a more rigorous footing using objective/quantitative classifications in future. 

\subsection{Robustness of Hubble Space}
\label{subsec:hubblespace}

The sensitivity of the Hubble space distribution to increasing
noise and decreasing resolution is shown in
Figure~\ref{fig:barvisibility}.  The top row of this figure
illustrates local Hubble space with plot symbols keyed to the bar
classifications from the RC3 (left panel), the {\em Revised
Shapley Ames Catalog} (middle panel), and visual inspection by van
den Bergh (right panel). A comparison of the local bar
classifications reveals a number of interesting trends. Firstly,
as noted earlier, a simple cut in this parameter space isolates
barred from unbarred spirals as classified in the RC3. A
comparison between the RC3 and RSA panels shows a generally
excellent agreement between the catalogs in the numbers of {\em
strongly} barred spirals. An independent visual inspection by Eskridge
et al.  (2000) also results in close agreement with the numbers in 
the RC3. Note however the lack of systems classed
as {\em weakly} barred in the RSA --- weakly barred systems in the
RC3 are generally classed as unbarred in the RSA, although
virtually all such objects lie above the dashed line in the
diagram (our proposed quantitative discriminator between barred
and unbarred galaxies). This is a rather pleasing graphical
demonstration of the fairly well-known fact (Binney \& Merrifield
1998) that the proportion of barred spirals in local catalogs
varies between $\sim 30$\% and $\sim60$\%, and highlights the
subjective nature of assigning concrete classifications to systems
in which a continuum of properties exists. Even more striking
evidence of this is the rightmost panel, which shows that van den
Bergh's criteria for classification as an SB galaxy are much
stricter than those of the RSA. As described in the previous
section, and is shown in Table~2, many systems classified as SB in
the RC3 and RSA are classified as weakly barred S(B) galaxies by
van den Bergh.

The middle row of Figure~\ref{fig:barvisibility} shows the Hubble
space distribution for galaxies degraded to the conditions in the
Northern Hubble Deep Field, with plot symbols color-coded
according to their {\em original} classifications in the local
images (ie. to the classifications shown in the top set of
panels). Over 90\% of galaxies originally classified as SB remain
above the dashed line, and these galaxies would thus also be
classified as strongly barred in the {\em Hubble Deep Field} using
the methodology of Abraham et al. (1999). While almost none of the
spirals classified as S(B) by van den Bergh would be misclassified
as regular spirals in the HDF, around half of the weakly barred
(SAB) systems from the RC3 have migrated to the unbarred portion
of the Hubble space diagram. These weakly barred objects would
have been classified as unbarred by the Abraham et al. (1999) HDF
study.  We conclude that, on the whole, the bars in strongly
barred spirals (i.e. those systems classified SB in the RSA and
RC3, and as SB or S(B) by van den Bergh in the present study),
would remain visible in galaxies at $z=0.7$ observed under the
conditions of the central {\em Hubble Deep Field}.

The situation for shallower observations is less sanguine.
The bottom row of Figure~\ref{fig:barvisibility} shows the Hubble
space distribution for \ii$<22$ mag galaxies degraded to the
signal-to-noise and resolution of the HDF Flanking Field
observations. It is clear that in data of this quality (ie.
two-orbit exposures) only around 40\% of strongly barred local
spirals (at $I_{814} < 22$ mag) would be classified as strongly
barred at z=0.7, in good agreement with the results obtained
by visual inspection summarized in Table~2.

\section{LUMINOSITY CLASSIFICATION AND GRAND DESIGN SPIRALS}
\label{sec:grand}

Luminosity classification of spiral galaxies (van den Bergh
1960abc) is intrinsically more challenging than is the assignment
of Hubble types. Because the fine details of spiral structure are
both smaller and of lower surface brightness than bars, assignment
of luminosity classes is also more difficult than bar-strength
classification. One would expect that only the strongest and most
extended spiral features will remain visible in degraded images,
i.e. only ``grand design'' spirals are expected to be recognizable
on noisy degraded images. Inspection of Table 1 shows that 27
local spirals have both luminosity classifications by Sandage and
luminosity classifications by van den Bergh in images degraded to
HDF quality. It is encouraging to see that these two sets of
luminosity classifications exhibit no statistically significant
systematic difference: the standard deviation of the
differences between the Sandage and van den Bergh classifications
is found to be only 0.6 luminosity classes. Not unexpectedly the
fraction (9/26 = 35\%) of spirals to which uncertain (marked as :)
luminosity classifications were assigned is greater for the
high-noise images in Table 2 than it is the fraction of uncertain
luminosity classifications (3/27 = 11\%) for the lower noise
images listed in Table 1. Since only three HDF spirals are known
to have 0.$60 < z < 0.80$ [see Table 2 of van den Bergh et al.
2000) it is not yet possible to establish if the fraction of grand
design spirals is lower in situ at $z \sim 0.7$ than it is for the
present sample of local spirals with images degraded to the
appearance that they would have at z = 0.70.

Since, as described earlier, luminosity classes cannot be assigned
to all spiral galaxies in which spiral structure is visible (e.g.
early-type spirals), and since arms may still be visible in
galaxies with insufficient signal to allow a reliable luminosity
classification, it is interesting to consider what fraction of
galaxies at $z=0.7$ would show spiral features of any sort. In an
attempt to determine this, one of us (RGA) visually inspected the
complete set of local and degraded images and assigned each galaxy
a numerical index (0=spiral structure is invisible, 1=hint of weak
structure, 2=obvious spirals structure) corresponding to
qualitative visibility of spiral features. These numbers are
included in Table~1, and Hubble space plots summarizing the change
in the visibility of spiral structure with redshift and noise are
shown in Figure~\ref{fig:spirclass}. In our local sample 97\% of
galaxies showed ``obvious'' spiral structure. Under the conditions
of the central HDF observations obvious spiral structure is still
seen in most (61\%) galaxies at $z=0.7$, with 82\% showing
at least a hint of structure. However, in the conditions
corresponding to the HDF flanking field observations only a
minority (33\%) of spiral galaxies now show obvious spiral
structure, with 55\% showing at least a hint of structure. Once
again, we conclude that the deep central HDF observations allow
quite robust conclusions to be drawn regarding galaxy morphology
at $z=0.7$, but that detailed morphological classifications (i.e.
those more detailed than crude Hubble types) in the Flanking
Fields should be treated with caution.

\section{DISCUSSION}
\label{sec:discussion}

It is now commonly accepted that a large fraction of the galaxy
population seen at large redshifts does not appear to fit
comfortably with the tuning fork classification scheme that
provides a satisfactory framework for the classification of
low-redshift galaxies in the $B$-band. While it seems likely that
many of these peculiar galaxies are not isolated disk systems, the
present paper does shed some light on the subset of high-redshift
spirals classified as peculiar. (Since the BSGS sample is selected from galaxies
with Hubble stages in the range $0\le T\le 9$, it does not include
systems classed as irregular or unclassifiable, and therefore the
BSGS is not suitable for quantifying the proportion of non-spiral
peculiars in deep images). Among the 101 BSGC galaxies, which
are located at $z \sim 0.0$, 12 (12\%) are classified as being
peculiar on images degraded to $z = 0.7$ with HDF resolution and
signal-to- noise ratio. For comparison, the data by van den Bergh
et al. (2000) for HDF + FF spiral/disk galaxies at $0.60 < z < 0.89$ show that
17 out of 37 (46\%) are classified as peculiar. In other words
approximately half of all spiral galaxies at look-back times of $\sim8$
Gyr do not fit well into the Hubble classification system.
Furthermore, van den Bergh et al. (2000) shows that early-type
galaxies are less likely to be peculiar than is the case for
objects of later type. Only 1/22 (5\%) of E-Sa-Sab galaxies are
classified as peculiar, whereas 11/16 (69\%) of Sb-Sbc galaxies
are so classified. The majority of ``Sc'' galaxies at $z \sim 0.7$
are so peculiar that many of them have probably been called
proto-Sc, Pec or ``?". In other words it looks like most compact
early-type galaxies started to approach their ``normal"
present-day morphology faster than was the case for more
extended objects of later morphological types.  A caveat is,
however, that it might be more difficult to recognize
peculiarities in compact images of early-type galaxies than it
would be to see such anomalies in the more open structure of
late-type galaxies.

\begin{figure*}[htb]
    \centering
    \includegraphics[width=4.5in,angle=90]{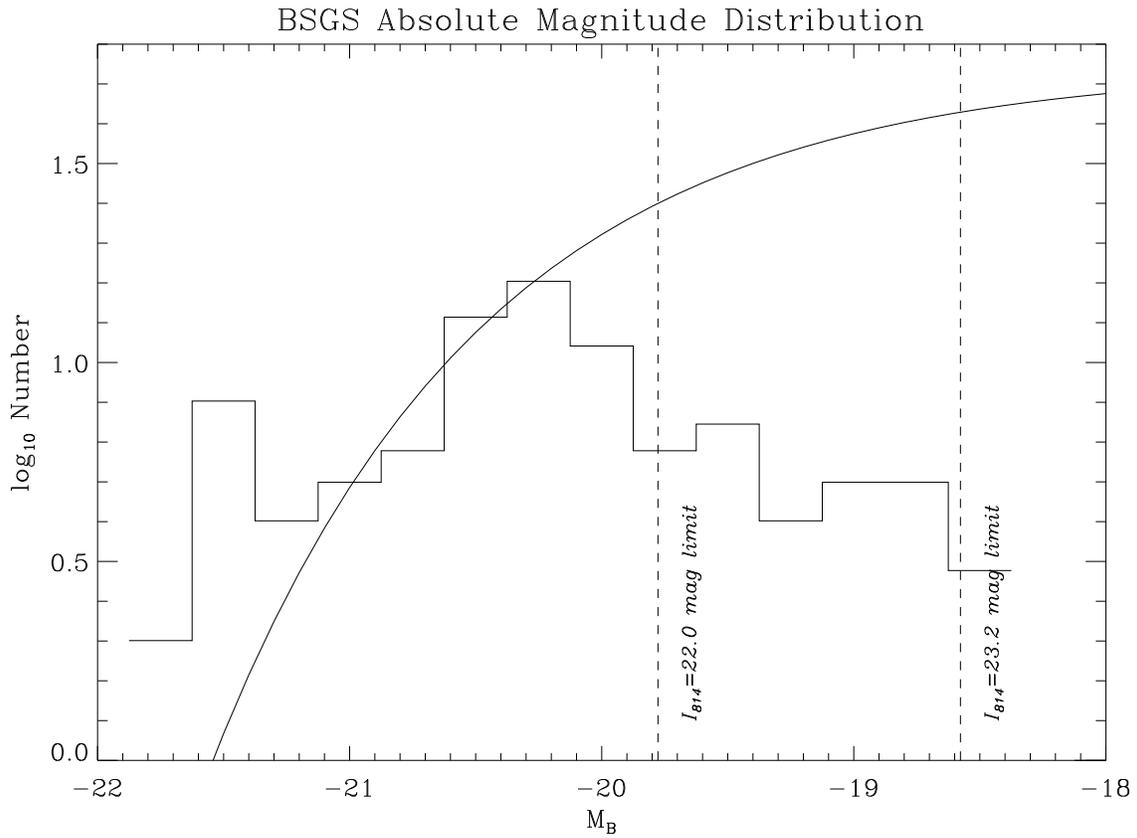}
    \caption{The absolute magnitude distribution of the BSGS. For a galaxy
    population at z=0.7, dashed vertical
    lines represent the absolute magnitude limits corresponding to $I_{F814W}=22$
    mag and $I_{F814W}=23.2$ mag for zero evolution.
    The curve represents an $\alpha=-1$ Schechter function normalized
    to the bright shoulder of the luminosity distribution. See text
    for further details.}
    \label{fig:absmag}
\end{figure*}

\begin{figure*}[htb]
    \centering
    \includegraphics[width=5.3in,angle=90]{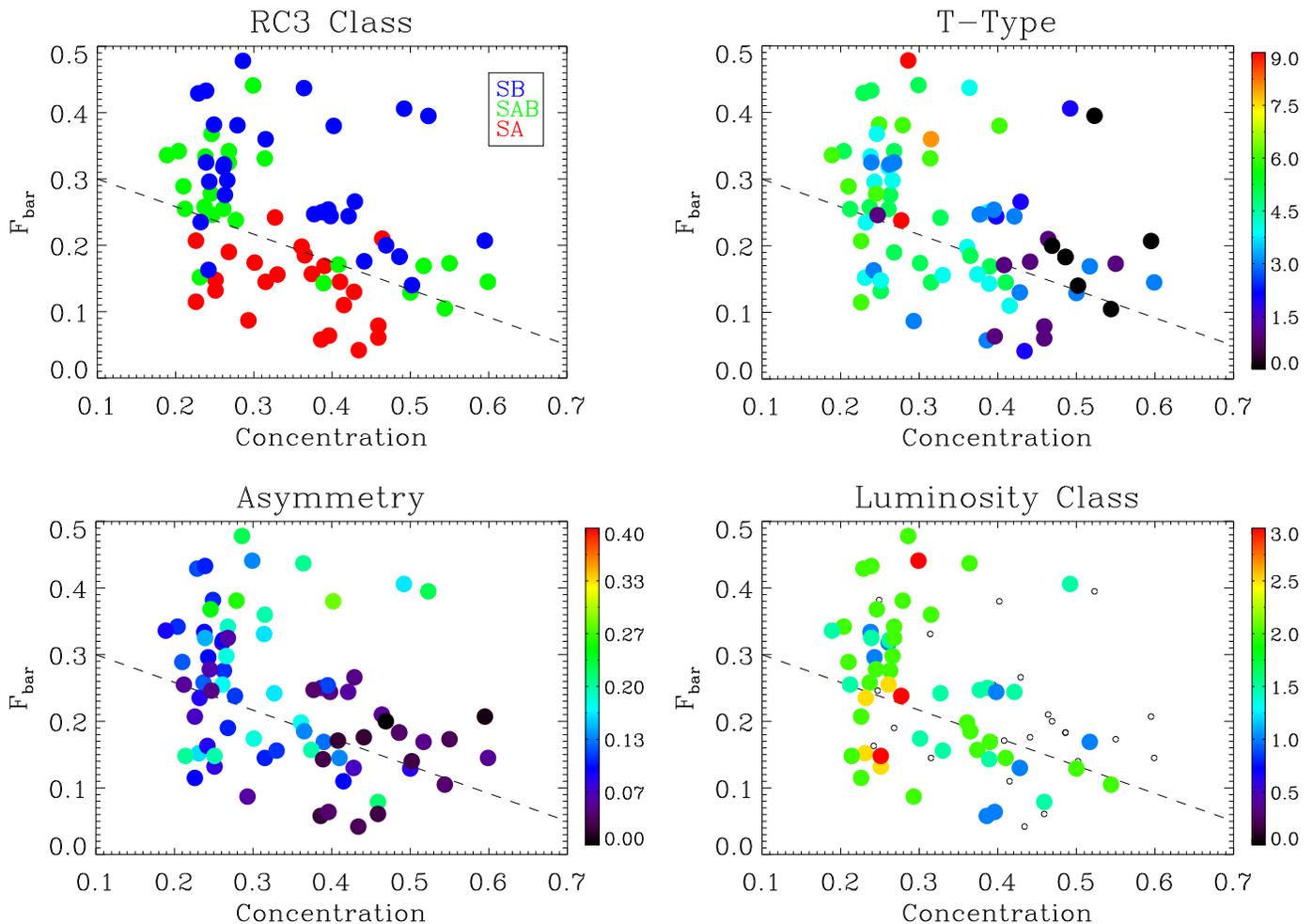}
    \caption{The distribution of our sample in
    the two-dimensional morphological parameter
    space defined in Abraham \& Merrifield (2000). [Top Left] Colored
    symbols subdivide the galaxy population into morphological
    ``form families'', based on classifications from the RC3.
    Unbarred (SA) spirals are shown in red, weakly barred spirals
    (SAB) are shown in green, and strongly barred (SB) spirals are
    shown in blue. The dashed line subdivides barred from unbarred
    galaxies rather cleanly. Unlike the corresponding figure for
    the Frei et al. (1996) sample (Abraham \& Merrifield 2000), no obvious
    ``tuning fork'' shape (or even bimodality) emerges from this
    diagram. At best a hint of bimodality appears when the diagram
    is restricted to symmetric galaxies, as described in the text.
    However, as was found in the analysis of the Frei et al (1996) sample,
    weakly barred spirals are predominantly late-type (ie. low
    central concentration). [Top Right] As for the previous panel,
    except with symbols keyed to Hubble stage via RC3 T-types. (At the
    extremes of the distribution, T=0
    corresponds to S0/a, T=9 corresponds to Sm. The full mapping between
    Hubble stage and T-type is shown at the top of Figure~\ref{fig:rc3comparison}).
    Note the evident
    gradient in Hubble stage with central concentration. [Lower
    Left] As for the previous panel, except with plot symbols
    keyed to rotational asymmetry, as defined in Abraham et al. (1996).
    Note how SB galaxies appear to
    be made up of two populations: a tight sequence of low
    asymmetry galaxies and an outlier population of highly
    asymmetric galaxies. [Lower Right] As for the previous panel,
    except with plot symbols keyed to luminosity class from the
    Revised Shapley Ames catalog. For convenience, luminosity
    classes have been mapped onto a numerical sequence (eg. type I
    = 1.0, II-III = 2.5, etc.). Galaxies with no luminosity
    classifications have been designated
    as class 0, and are shown as small black open circles.}
\label{fig:local}
\end{figure*}

\begin{figure*}[htb]
  \centering
  \includegraphics[width=7.0in]{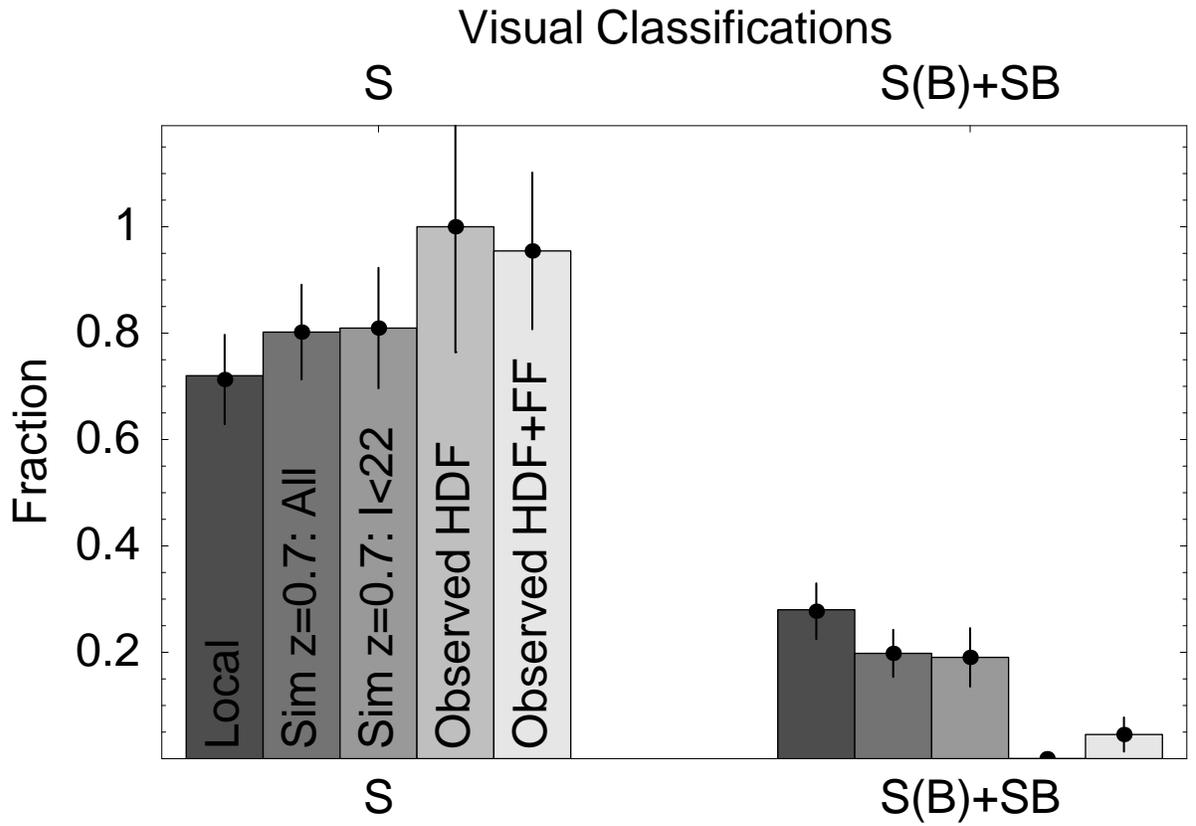}
  \caption{Summary of classification results based on visual inspection by
van den Bergh. Histograms are
divided into two groups, showing systems classed as unbarred (left) and
barred (right).  Within these categories results are shown from left to
right corresponding to (1) un-degraded images, (2) simulated HDF images, (3)
simulated 2-3 orbit HST imaging for objects brighter than I=22 mag, (4) observed data from the Hubble Deep Field, (5) Observed data from the study of van den Bergh et al. (2000), corresponding to the HDF + shallow flanking field images. } 
\label{fig:secsummary}
\end{figure*}

\begin{figure*}[htb]
  \centering
  \includegraphics[width=5.4in,angle=90]{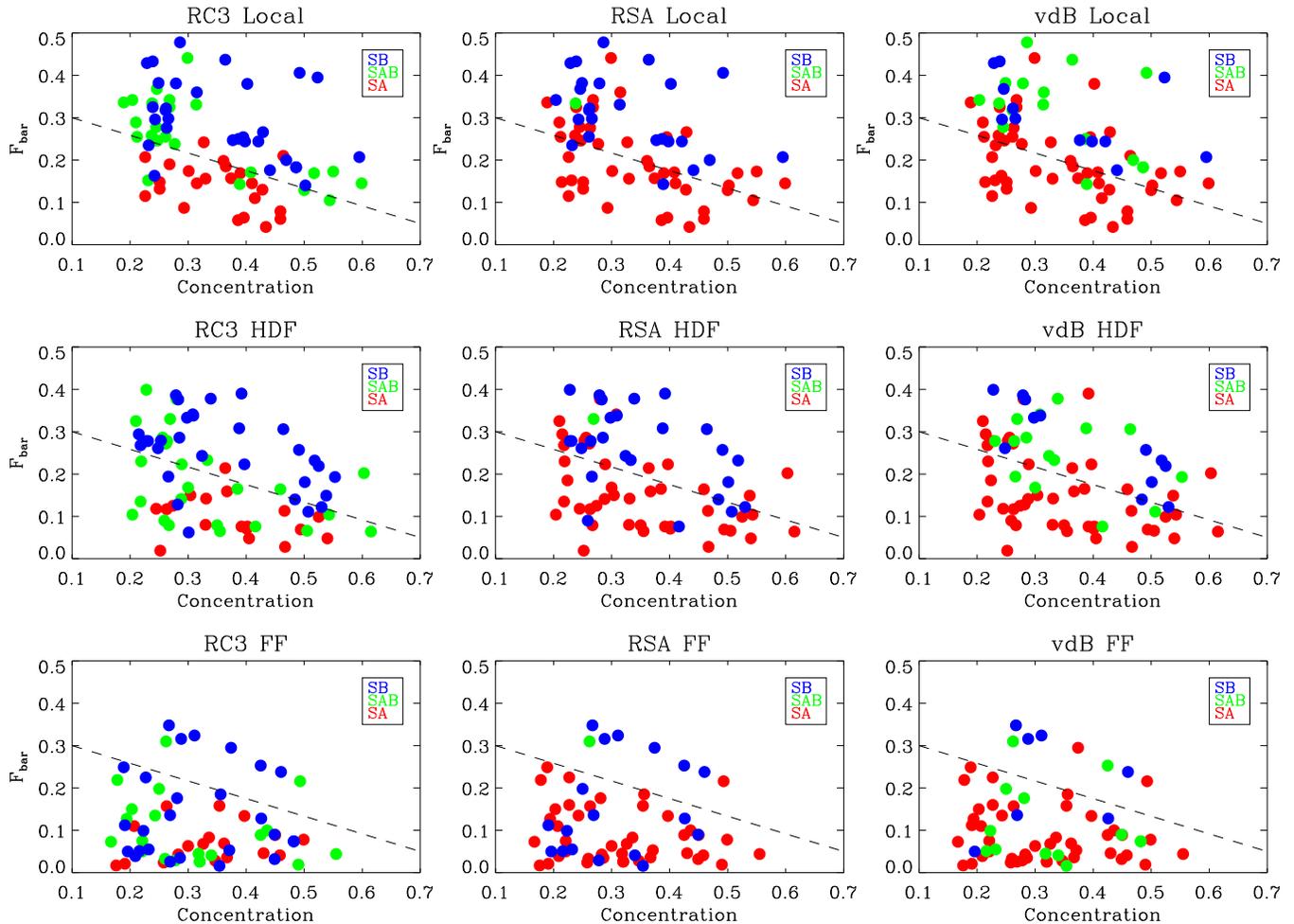}
  \caption{An illustration of the degradation of bar visibility
  with redshift and noise, and of observer-to-observer agreement
  in visual bar classifications. [Top row] The local Hubble space
  diagram with plot symbols keyed to the bar classifications from
  the RC3 (left panel), the RSA (middle panel), and visual
  inspection by van den Bergh (right panel), as described in the
  text. Note the excellent agreement between the bar
  classifications in the RC3 and position above the dashed line in
  Hubble space. The middle panel illustrates the well-known fact
  that the RSA adopts stricter criteria for classification of
  barred spirals than the RC3 (Binney \& Merrifield 1998). The
  rightmost panel shows that van den Bergh's visual
  classifications are closer to the RSA than to the RC3, although
  van den Bergh classifications allow for more dynamic range
  within the barred spiral category: many objects classified as
  S(B) by van den Bergh are classified as SB in the RSA. [Middle
  Row] As for the top row, except showing the Hubble space diagram
  for the sample degraded to the conditions in the Northern Hubble
  Deep Field. Colors are keyed to {\em local} classifications.
  Strongly barred galaxies are clearly still visible in these
  data, though weakly barred systems have now migrated to the
  unbarred portion of the Hubble space diagram. [Bottom Row] As
  for the previous row, except showing the Hubble space diagram
  for the sample degraded to the conditions of the HDF Flanking
  Field observations. Only a few strongly barred spiral galaxies
  are evident in data of this quality. } \label{fig:barvisibility}
\end{figure*}

\begin{figure*}[htb]
  \centering
  \includegraphics[width=5.3in,angle=90]{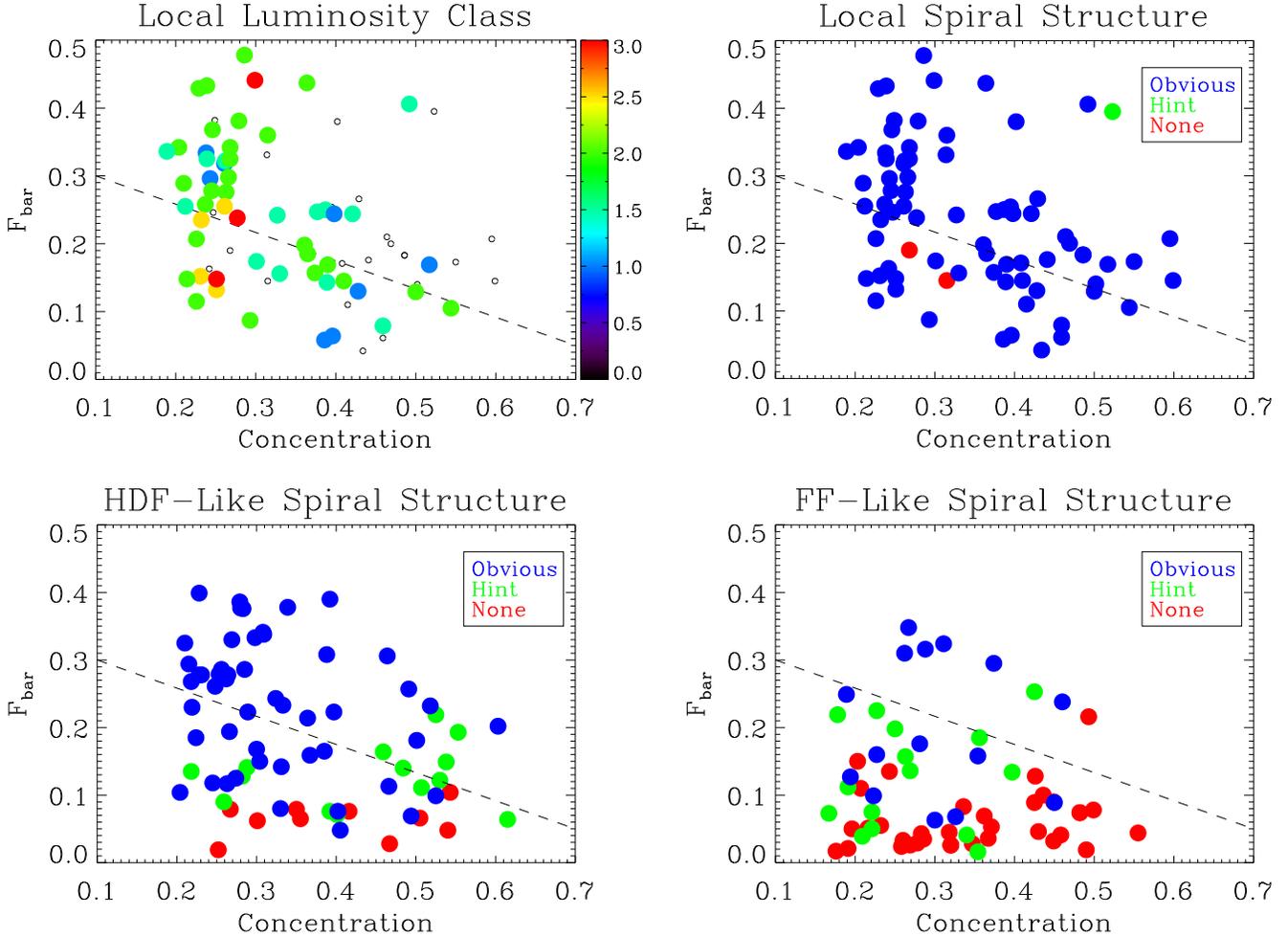}
  \caption{A montage illustrating the degradation of spiral structure
  visibility with redshift and noise. [Top left] Local luminosity classifications.
  Objects with no luminosity classifications are shown as small open
  circles. [Top right] Visually assessed visibility of of spiral structure
  keyed to symbol color. Galaxies with obvious spiral structure are
  shown in blue, galaxies with a hint of spiral structure are shown
  in green, and galaxies no visible spiral structure are shown
  in red. As described in the text, many objects for
  which no local luminosity classification is possible still show obvious
  spiral structure. Most of these galaxies are early-type systems
  (high central concentration).
  [Lower left] The corresponding plot for galaxies at a synthetic
  redshift of z=0.7 as seen under the conditions of the central HDF
  field. Colors are keyed to classifications made directly
  from the synthetic high-redshift
  galaxies. Obvious spiral structure is still seen in most
  galaxies. [Lower right] As for the previous plot, except for conditions
  corresponding to the HDF flanking field observations. Galaxies
  showing obvious spiral structure are now a minority.}
\label{fig:spirclass}
\end{figure*}

Our numerical experiments have given us some insight into the
proportion of normal spirals which might be misclassified as
peculiar due to noise and sampling. However, on the basis of
Figure~\ref{fig:absmag}, we are more cautious about using these
simulations to say very much about the relative space densities of
different classes of spirals seen in deep images. It might be
argued that a luminosity function peaked at $M^\star$ is a generic
consequence of sampling a Shecheter function affected by Malmquist
bias, so the overall shape of the luminosity distribution for the
local galaxies in Figure~\ref{fig:absmag} is also generically
similar to what one also expects from a deep HST imaging campaign
(in the absence of strong field galaxy evolution). However, the
shape of this distribution is largely due to the absence of
intrinsically faint galaxies at high redshifts, and the
distribution of absolute magnitudes within a narrow redshift shell
may look quite different from the absolute magnitude distribution
for the sample as a whole. Since morphology is a strong function
of rest wavelength, it is likely that any morphology-based
campaign would isolate galaxies in redshift shells in order to
construct fair samples to compare with local calibration data (eg.
van den Bergh 2000).  Clearly then, for the specific purpose of
calibrating the mix of different types of spirals fixed to a
narrow redshift shell, an ideal local comparison sample should be
drawn from an enormous parent population using a Schechter
function-shaped distribution. We hope the sheer number of galaxies
imaged by the {\em Sloan Digital Sky Survey} will soon make this
possible, and at that point an extensive comparison between low-z
and high-z Hubble space distributions will become straightforward.

In the meantime, we emphasize that our goals in the present paper
are much more limited. Our main goal is to determine the extent to
which a typical local spiral galaxy is likely to be misclassified
at high redshift (or, equivalently, how far a galaxy situated at a
particular position in local Hubble space is likely to shift by
$z=0.7$). For this purpose, Figure~\ref{fig:absmag} is less
important than Figure~\ref{fig:rc3comparison}, and the
incompleteness of our BSGS sample at the faint end relative to the
numbers expected from sampling a Schechter function-shaped
distribution is not important so long as there are {\em enough}
low-luminosity galaxies with a mix of morphological types in our
sample to allow us to gauge how much a typical low-luminosity
spiral is likely to move in a Hubble space diagram as a function
of resolution and signal-to-noise. For example, since there are 28
galaxies fainter than $M_B=-19.5$ in the subset of the BSGS
analyzed in this paper, and since $\sim 2/3$ of the SB subset of
these would be classified SB at $z=0.7$ in the HDF, we think it
very unlikely that either low signal-to-noise or resolution
effects can explain the apparent absence of barred spirals in the
HDF data.

Even after the numerical experiments described in this paper we do
remain concerned that cosmic variance might be the true
explanation of the perceived absence of barred spirals at high
redshifts. There is no known environmental dependence of bar strength on local galaxy density, though this possible dependence has not yet been definitively explored, and this is clearly a topic worthy of future study. It 
is conceivable that spirals in clusters might exhibit quite different distributions
bar strengths due to numerous factors which might heat disks and supress bar instabilities, such as tidal harrassment (Moore et al. 1996) or gas stripping (Abraham et al. 1996).
Furthremore, the volume encompassed to $z=1$ in HDF-like observation
is tiny. Assuming the non-evolving local luminosity function given
by Gardner et al. (1997), only around 30 $L^\star$ galaxies (of all 
morphological types,
and with a range of inclinations) are expected
to be visible (and, in fact, about this number is seen) in the
redshift range $0<z<1$ in an HDF image. About 10 of these should
be in the redshift range $0.6<z<0.8$. (In fact, only 3 spirals in
this redshift range are seen in the Northern HDF). Even after
adding together both Hubble Deep Fields, the numbers remain
appallingly low for addressing a topic as important as the
possible disappearance of bars; clearly more data are needed in
order to conclusively rule out the possibility that cosmic
variance could explain the absence of barred spirals in the Hubble
Deep Fields .

\section{CONCLUSIONS}
\label{sec:conc}.

In this paper we have studied the effects of image degradation on
galaxy classification by (1) lowering resolution and by (2)
increasing noise. Images were studied at both the noise level of
the deep Hubble Deep Field images and at that of the shallower
Flanking Field images. From comparison of the images of an
unbiased sample of local (z = 0.0) galaxies, with those of the
same galaxies degraded to the appearance that they would have at z
= 0.7, it is found that about $\sim 2/3$ of all SB galaxies are still
classified as SB or S(B) at the resolution and noise level of the
Hubble Deep Field, dropping to around 50\% in the flanking fields.
Image degradation can account for some, but not all, of the
observed (van den Bergh et al. 1996, Abraham et al. 1999) decrease
in the fraction of SB galaxies with increasing redshift. Since
luminosity classification is more challenging than determination
of Hubble type it is not surprising that only a fraction of
the degraded images could be assigned DDO luminosity
classifications. It is, however, of interest to note that the
luminosity classifications that could be made on degraded images
agree well with those made on the original full- resolution
noise-free images. We conclude that the fraction of grand design 
spirals appears to decrease with increasing
redshift, although the number of galaxies observed at suitable
signal-to-noise levels remains small and more observations
are needed to better establish the demise of grand design
spirals with redshift.

On the basis of our simulations, it is clear that when local
spiral galaxies are viewed at comparable resolution and noise
levels to the HDF data, these systems are not systematically
misclassified as peculiar. This lends credence to the notion that
the fraction of peculiar disk galaxies is dramatically higher at
$z\sim0.7$ than it is at $z =0.0$. In the study by van den Bergh
et al. (2000) for HDF + FF galaxies at $0.60 < z < 0.80$, 17 out
of 37 (46\%) spirals are classified as peculiar. In other words
approximately half of all disk galaxies at look-back times of
$\sim8$ Gyr do not fit well into the Hubble classification system.
This can be compared to the only 12\% of artificially redshifted
Sb+Sbc+Sc galaxies classified as peculiar in the present sample
viewed at the same resolution and noise level. Among early-type
spirals, the fraction of peculiars increases more slowly with
redshift. These results suggest that (1) the fraction of
intrinsically peculiar disk galaxies grows dramatically with
increasing look- back time, and (2) the fraction of peculiar
galaxies at $z\sim0.7$ is much larger among late-type galaxies
than it is among early-type spirals. It seems that late-type
spirals approach their ``normal'' morphology more slowly than do
galaxies of early type.


\vfill\eject

\begin{deluxetable}{lcccccccc}
\tablecolumns{9}\tablecaption{Summary of Visual Classifications}
\tabletypesize{\footnotesize}
\tablewidth{0pc}\tablehead{
  \colhead{ID} &
  \colhead{RC3\tablenotemark{(1)}} &
  \colhead{RSA\tablenotemark{(2)}} &
  \colhead{z=0.0\tablenotemark{(3)}} &
  \colhead{z=0.7 HDF\tablenotemark{(3)}} &
  \colhead{z=0.7 FF\tablenotemark{(3)}} &
  \colhead{$M_{814}^{z=0.7}$\tablenotemark{(4)}} &
  \colhead{$ 1 - {b\over a}$\tablenotemark{(5)}} &
  \colhead{Comment}
}
\startdata
NGC0150 & .SBT3*. & Sbc p II & Sbc II [2] & Sc [2] & \nodata & 22.29 & 0.47 &  \\
NGC0157 & .SXT4.. & Sc II-III  & Sc I-II [2] & Sc II [2] & Sc I [2] & 20.85 & 0.25 &  \\
NGC0210 & .SXS3.. & Sb I & Sb I [2] & S(B?)b II [2] & Sb [0] & 21.58 & 0.12 & Lens \\
NGC0278 & .SXT3.. & Sbc II & Sbc: [2] & Sa: [0] & \nodata & 23.53 & 0.06 &  \\
NGC0289 & .SBT4.. & SBbc I-II  & S(B)b I [2] & S(B?)b II [2] & Sb [0] & 21.72 & 0.44 &  \\
NGC0428 & .SXS9.. & Sc III & Sbc III: [2] & Sb: p [2] & \nodata & 22.89 & 0.09 &  \\
NGC0488 & .SAR3.. & Sab I & Sa I [2] & Sa: [2] & Sab [0] & 20.51 & 0.26 &  \\
NGC0578 & .SXT5.. & Sc II & Sbc I-II [2] & Sbc II [2] & Sbc [1] & 21.53 & 0.47 &  \\
NGC0613 & .SBT4.. & SBb II & S(B)b I [2] & S(B)b I [2] & SBb II: [1] & 20.90 & 0.44 &  \\
NGC0625 & .SBS9\$/ & S p & Amorph. [0] & Sab: [0] & \nodata & 24.89 & 0.47 & Dusty \\
NGC0685 & .SXR5.. & SBb II  & S(B)c III: [2] & S(B)c: [2] & Sc: II: [1] & 21.92 & 0.19 &  \\
NGC0779 & .SXR3.. & Sb I-II  & Sab II: [2] & Sb [0] & \nodata & 22.22 & 0.66 &  \\
NGC0908 & .SAS5.. & Sc I-II & Sc I-II [2] & Sc II [2] & Sc II [2] & 20.70 & 0.57 &  \\
NGC1042 & .SXT6.. & Sc I-II  & Sc II [2] & Sc II [2] & Sc [1] & 21.90 & 0.19 &  \\
NGC1058 & .SAT5.. & Sc II-III & Sc [2] & Sa [0] & \nodata & 24.60 & 0.03 &  \\
NGC1073 & .SBT5.. & SBc II & SBbc II [2] & SBb I [2] & \nodata & 22.19 & 0.07 &  \\
NGC1084 & .SAS5.. & Sc II & Sc II: [2] & Sbc [2] & Sb [0] & 21.57 & 0.42 &  \\
NGC1087 & .SXT5.. & Sc III-IV & Sc p? [2] & Sbc: p [2] & Sbc [0] & 21.62 & 0.40 &  \\
NGC1187 & .SBR5.. & Sbc II & Sc I-II [2] & S(B?)c: II [2] & Sc II [1] & 21.47 & 0.27 &  \\
NGC1300 & .SBT4.. & SBb I  & SBb I [2] & SBb I [2] & SBb I [2] & 21.14 & 0.40 &  \\
NGC1302 & RSBR0.. & Sa & Sa I: [2] & S p [1] & Sb: [0] & 21.55 & 0.07 & Outer shells? \\
NGC1309 & .SAS4*. & Sc II & Sc III: [2] & Sc: [2] & Sc II: [1] & 21.29 & 0.04 &  \\
NGC1317 & .SXR1.. & Sa & Sa [2] & Sab: [0] & Sb: [0] & 21.71 & 0.02 &  \\
NGC1350 & PSBR2.. & Sa   & Sb I: [2] & S(B:)b [2] & Sb II: [2] & 20.88 & 0.58 &  \\
NGC1371 & .SXT1.. & Sa & Sab I [2] & Sb: [1] & Sb [0] & 21.84 & 0.28 &  \\
NGC1385 & .SBS6.. & SBc & Sc [2] & Sc: p [2] & Sc p [2] & 21.00 & 0.34 & Bar-like knots \\
NGC1493 & .SBR6.. & SBc & S(B)bc II-III [2] & S(B)b: [2] & \nodata & 22.80 & 0.09 &  \\
NGC1511 & .SA.1P* & Sc p & S [2] & Sb: p [1] & \nodata & 23.39 & 0.60 & Dusty, edge-on \\
NGC1559 & .SBS6.. & SBc II  & S(B:)c [2] & Sc [2] & Sc [2] & 21.53 & 0.47 &  \\
NGC1617 & .SBS1.. & Sa & Sab II: [2] & Sab [0] & \nodata & 22.44 & 0.51 &  \\
NGC1637 & .SXT5.. & SBc II-III & Sbc II [2] & S [1] & \nodata & 23.27 & 0.21 & Asymmetric \\
NGC1703 & .SBR3.. & ... & Sc I-II [2] & Sbc: [1] & \nodata & 22.11 & 0.13 &  \\
NGC1792 & .SAT4.. & Sc II & Sc [2] & Sc [2] & Sc [0] & 21.49 & 0.52 &  \\
NGC1808 & RSXS1.. & Sbc p & S p [2] & Sc: [2] & Sc: [0] & 21.82 & 0.59 & Dusty \\
NGC1832 & .SBR4.. & SBb I & S(B)b: II [2] & Sc I: [2] & Sc II [1] & 21.75 & 0.31 &  \\
NGC2090 & .SAT5.. & Sc II & Sbc II [2] & Sab [0] & \nodata & 23.34 & 0.57 &  \\
NGC2139 & .SXT6.. & SBc & S(B:)c (t?) [2] & Sc: t: [2] & S(B?)c [1] & 21.72 & 0.07 &  \\
NGC2196 & PSAS1.. & Sab I & Sb II [2] & Sb I [2] & Sb II [2] & 21.04 & 0.17 &  \\
NGC2442 & .SXS4P. & SBbc II & SBbc I-II [2] & SBb p [2] & S(B?)bc II [2] & 21.62 & 0.33 &  \\
NGC2775 & .SAR2.. & Sa & Sa [2] & Sab: [0] & Sab [0] & 21.45 & 0.18 &  \\
NGC3166 & .SXT0.. & Sa & Sab [2] & Sab: [0] & Sa: [0] & 21.77 & 0.55 &  \\
NGC3169 & .SAS1P. & Sb t I-II & Sb p (t?) [2] & Sb p? [2] & Sb p [0] & 21.66 & 0.40 & Dusty \\
NGC3223 & .SAS3.. & Sb II & Sbc I-II [2] & Sbc [2] & Sbc II [2] & 20.57 & 0.31 &  \\
NGC3275 & .SBR2.. & SBab I & SBb II [2] & Sb t? [2] & Sbc I: [2] & 20.35 & 0.09 &  \\
NGC3338 & .SAS5.. & Sbc I-I & Sbc I [2] & Sbc II: [2] & \nodata & 22.27 & 0.59 & Asymmetric \\
NGC3423 & .SAS6.. & Sc II & Sc III [2] & Sbc [2] & \nodata & 23.03 & 0.19 &  \\
NGC3511 & .SAS5.. & Sc III & Sc p [2] & Sc: [2] & \nodata & 22.31 & 0.66 & Dusty \\
NGC3513 & .SBT5.. & SBc II & SBbc III: [2] & S(B)c [2] & \nodata & 22.62 & 0.21 &  \\
NGC3583 & .SBS3.. & Sbc & Sbc II [2] & Sbc p [2] & Sbc p [1] & 21.57 & 0.39 &  \\
NGC3593 & .SAS0*. & Sa p & Sb III: [0] & Sa [0] & \nodata & 24.12 & 0.51 & Edge-on \\
NGC3646 & .RING.. & Sbc p II & Sc p I [2] & Sc I: [2] & Sc p I [2] & 20.00 & 0.30 &  \\
NGC3675 & .SAS3.. & Sb II & Sab II: [2] & Sab [1] & \nodata & 22.76 & 0.51 &  \\
NGC3681 & .SXR4.. & SBb I-II & S(B)ab II: [2] & Sab [0] & \nodata & 22.37 & 0.06 &  \\
NGC3726 & .SXR5.. & Sbc II & Sc II: [2] & Sc [1] & \nodata & 22.08 & 0.38 &  \\
NGC3810 & .SAT5.. & Sc II & Sc I-II [2] & Sbc II [1] & \nodata & 22.42 & 0.30 &  \\
NGC3877 & .SAS5*. & Sc II & Sb III: [2] & Sb: [0] & \nodata & 23.24 & 0.78 & Edge-on \\
NGC3885 & .SAS0.. & Sa & Sb: [2] & Sb [1] & S(B?)b: [0] & 21.55 & 0.65 & Lens? \\
NGC3887 & .SBR4.. & SBbc II-III & Sbc III [2] & Sc [2] & \nodata & 22.05 & 0.31 &  \\
NGC3893 & .SXT5*. & Sc I & Sbc II: [2] & Sb: [2] & \nodata & 22.54 & 0.43 & Asymmetric \\
NGC4027 & .SBS8.. & Sc II & S(B)bc? p [2] & S p [2] & Sbc: p [2] & 21.69 & 0.27 & Tidal arm? \\
NGC4030 & .SAS4.. & ... & Sbc I-II [2] & Sbc [2] & Sb [0] & 21.68 & 0.23 &  \\
NGC4136 & .SXR5.. & Sc I-II & Sbc II-III: [2] & Sa [0] & \nodata & 24.70 & 0.09 &  \\
NGC4254 & .SAS5.. & Sc I-II & Sc I-II [2] & Sc I [2] & Sc I [2] & 19.48 & 0.26 &  \\
NGC4314 & .SBT1.. & SBa p & SB0/SBa p [2] & S(B?)b? [1] & \nodata & 22.78 & 0.47 & Edge-on \\
NGC4414 & .SAT5\$. & Sc II & Sbc II-III: [2] & Sb: [1] & \nodata & 22.77 & 0.45 &  \\
NGC4457 & RSXS0.. & Sb II  & Sab [2] & Sab: [0] & \nodata & 23.49 & 0.30 &  \\
NGC4487 & .SXT6.. & Sc II & S(B?)c III [2] & Sc [2] & \nodata & 22.51 & 0.36 &  \\
NGC4504 & .SAS6.. & Sc II & Sc III [2] & Sbc: [2] & \nodata & 23.04 & 0.35 &  \\
NGC4580 & .SXT1P. & Sc & Sb II: [2] & Sbc [1] & \nodata & 22.40 & 0.24 &  \\
NGC4593 & RSBT3.. & SBb I-II & SBb I-II [2] & SBb II [2] & SBb [2] & 20.70 & 0.42 &  \\
NGC4618 & .SBT9.. & SBbc II & S(B)bc III-IV [2] & S p [2] & \nodata & 23.57 & 0.28 & Asymmetric \\
NGC4643 & .SBT0.. & SB0/SBa & SBa [2] & Sb: [1] & \nodata & 22.01 & 0.22 & Edge-on? \\
NGC4665 & .SBS0.. & ... & S(B)a [2] & Sb: [1] & \nodata & 22.08 & 0.11 &  \\
NGC4689 & .SAT4.. & Sc II-III & Sbc II? [2] & Sc: [2] & Sbc p [1] &  & 0.00 & Asymmetric \\
NGC4691 & RSBS0P. & Amorph. & SB p [1] & Sab: [1] & \nodata & 22.54 & 0.33 &  \\
NGC4699 & .SXT3.. & Sab/Sa & Sa [2] & Sab [1] & Sab [0] & 20.57 & 0.29 &  \\
NGC4772 & .SAS1.. & ... & Sab [2] & Sa [0] & \nodata & 23.10 & 0.50 &  \\
NGC4781 & .SBT7.. & Sc III & Sc III? [2] & Sb: [1] & \nodata & 23.06 & 0.54 &  \\
NGC4856 & .SBS0.. & S0/Sa & SBa/S0 [1] & S(B?)b [0] & \nodata & 22.07 & 0.63 &  \\
NGC4902 & .SBR3.. & SBb I-II & SBb I [2] & SBbc I [2] & SBbc I [2] & 20.46 & 0.08 &  \\
NGC4941 & RSXR2*. & Sab II & Sb [2] & Sb [0] & \nodata & 23.34 & 0.47 &  \\
NGC4995 & .SXT3.. & Sbc II & Sbc II: [2] & Sbc [2] & \nodata & 22.26 & 0.35 &  \\
NGC5005 & .SXT4.. & Sb II & Sb III [2] & Sb [0] & S(B?)b [0] & 21.73 & 0.54 &  \\
NGC5121 & PSAS1.. & Sa  & Sa [2] & Sb: [0] & Sbc: [0] & 21.68 & 0.17 &  \\
NGC5161 & .SAS5*. & Sc I & Sbc(t?) I-II [2] & Sbc II [2] & Sc: p? [2] & 21.24 & 0.61 &  \\
NGC5371 & .SXT4.. & Sb/SBb I & S(B?)bc I [2] & Sc I [2] & Sbc p I [2] & 20.35 & 0.42 &  \\
NGC5448 & RSXR1.. & Sa & Sb II: [2] & Sbc: [2] & Sb: [2] & 21.84 & 0.72 &  \\
NGC5676 & .SAT4.. & Sc II & Sbc p II [2] & S p [2] & Sc [1] & 21.27 & 0.57 & Asymmetric \\
NGC5701 & RSBT0.. & SBa & S(B)ab I: [2] & S(B?)b: [1] & \nodata & 22.02 & 0.21 &  \\
NGC5850 & .SBR3.. & SBb I-II & SBb II [2] & SBb II [2] & SBb II [2] & 20.68 & 0.39 &  \\
NGC6753 & RSAR3.. & Sb I & Sb II [2] & Sb II [2] & Sb II: [1] & 20.53 & 0.13 &  \\
NGC6907 & .SBS4.. & SBbc II & SBbc I-II [2] & SBbc II [2] & S(B)bc I [2] & 20.51 & 0.15 &  \\
NGC7083 & .SAS4.. & Sb I-II & Sbc I-II [2] & Sbc II [2] & Sbc II [2] & 20.49 & 0.46 &  \\
NGC7184 & .SBR5.. & Sb II & Sb III: [2] & Sb [2] & Sbc [2] & 20.56 & 0.73 &  \\
NGC7205 & .SAS4.. & Sb III & Sbc [2] & Sb p [2] & Sc p [1] & 21.72 & 0.50 & Asymmetric \\
NGC7412 & .SBS3.. & Sc I-II & Sc I [2] & Sbc II [2] & Sc I [2] & 21.81 & 0.23 &  \\
NGC7418 & .SXT6.. & Sc II   & Sc II [2] & Sc II [2] & Sc [1] & 21.90 & 0.26 &  \\
NGC7552 & PSBS2.. & SBb I-II & S(B)bc II [2] & S(B?)b [2] & S(B)b [1] & 21.22 & 0.44 &  \\
NGC7582 & PSBS2.. & SBab & S(B?)b II [2] & Sb: [2] & Sb p [0] & 21.52 & 0.58 & Lens \\
NGC7606 & .SAS3.. & Sb I & Sb I [2] & Sbc II [2] & Sbc II [0] & 20.93 & 0.58 &  \\
NGC7713 & .SBR7*. & Sc II-III & Sc III-IV [2] & Sb: [1] & \nodata & 23.28 & 0.51 &  \\
\enddata
\tablenotetext{(1)}{Classification code from Third Reference Catalog. The
standardized coding scheme from the RC3 is shown; note that weakly
barred spirals of type SAB are encoded as type SX.}
\tablenotetext{(2)}{Classification from Revised Shapley-Ames Catalog.}
\tablenotetext{(3)}{Visual classification by vdB.
Objects with $m_{814}>22$ mag in the flanking fields are excluded.
The number in square brackets following each classification indicates the qualitiative visibility of
spiral structure as assessed by RGA (0=invisible, 1=hint of structure, 2=obvious spiral structure).}
\tablenotetext{(4)}{Apparent F814W-band 2$\sigma$ isophotal magnitude of simulated $z=0.7$ HDF image.}
\tablenotetext{(5)}{1-axial ratio of simulated z=0.7 HDF image.}
\end{deluxetable}
\vfill\eject

\begin{deluxetable}{lcccccccc}
\tablecolumns{5}
\tablecaption{Frequency Distribution of Barred Spirals from Visual Classification}
\tabletypesize{\footnotesize} \tablewidth{0pc}\tablehead{
  \colhead{Type} &
  \colhead{Carnegie} &
  \colhead{z=0} &
  \colhead{z=0.7 HDF} &
  \colhead{z=0.7 FF}
} \startdata
S         & 69 (71\%)                      &  71   (70\%)  & 82 (81\%) &  47   (81\%) \\
S(B)      & 1   (1\%)\tablenotemark{1}     &  17   (17\%)  & 12 (12\%) &   6   (10\%) \\
SB        & 26 (26\%)                      &  11.5 (11\%)  &  7 (7\%)  &   5   (9\%) \\
S(B)+SB   & 26 (26\%)                      &  28.5 (28\%)  & 19 (19\%) &  11  (19\%) \\
\enddata
\tablenotetext{1}{Sandage \& Tammann (1991) do not use de Vaucouleurs
type type S(B).}
\end{deluxetable}

\end{document}